\documentclass[aps,pra,twocolumn,amsmath,amssymb,nofootinbib,superscriptaddress]{revtex4}

\newcommand{\bra}[1]{\langle#1|}
\newcommand{\ket}[1]{|#1\rangle}

\usepackage[pdftex]{graphicx}
\usepackage{mathrsfs}
\usepackage[usenames,dvipsnames]{color}
\usepackage[colorlinks]{hyperref}

\begin{document}

\bibliographystyle{apsrev}

%
%

\title{Sampling generalized cat states with linear optics is probably hard}

%
%

\author{Peter P. Rohde}
\email[]{dr.rohde@gmail.com}
\homepage{http://www.peterrohde.org}
\affiliation{Centre for Engineered Quantum Systems, Department of Physics \& Astronomy, Macquarie University, Sydney NSW 2113, Australia}

\author{Keith R. Motes}
\affiliation{Centre for Engineered Quantum Systems, Department of Physics \& Astronomy, Macquarie University, Sydney NSW 2113, Australia}

\author{Paul A. Knott}
\affiliation{School of Physics and Astronomy, University of Leeds, Leeds LS2 9JT, United Kingdom}
\affiliation{NTT Basic Research Laboratories, NTT Corporation, 3-1 Morinosato-Wakamiya, Atsugi, Kanagawa 243-0198, Japan}

\author{Joseph Fitzsimons}
\affiliation{Singapore University of Technology and Design, 20 Dover Drive, Singapore.}
\affiliation{Centre for Quantum Technologies, National University of Singapore, 3 Science Drive 2, Singapore.}

\author{William J. Munro}
\affiliation{NTT Basic Research Laboratories, NTT Corporation, 3-1 Morinosato-Wakamiya, Atsugi, Kanagawa 243-0198, Japan}

\author{Jonathan P. Dowling}
\affiliation{Hearne Institute for Theoretical Physics and Department of Physics \& Astronomy, Louisiana State University, Baton Rouge, LA 70803}

\date{\today}

\frenchspacing

%
%

\begin{abstract}
Boson-sampling has been presented as a simplified model for linear optical quantum computing. In the boson-sampling model, Fock states are passed through a linear optics network and sampled via number-resolved photodetection. It has been shown that this sampling problem likely cannot be efficiently classically simulated. This raises the question as to whether there are other quantum states of light for which the equivalent sampling problem is also computationally hard. We present evidence, without using a full complexity proof, that a very broad class of quantum states of light --- arbitrary superpositions of two or more coherent states --- when evolved via passive linear optics and sampled with number-resolved photodetection, likely implements a classically hard sampling problem.
\end{abstract}

\maketitle

%
%

\section{Introduction}

Linear optical quantum computing (LOQC) \cite{bib:KLM01, bib:kok2007QC} has become a leading candidate for the implementation of large-scale universal quantum computation \cite{bib:NielsenChuang00}. While LOQC is possible in principle, the technological requirements are daunting, requiring technologies that are not readily available today, such as fast feed-forward and optical quantum memory. Thus, the search for simplified, more technologically realistic models for LOQC is a priority. 

One recent proposal, by Aaronson \& Arkhipov (AA) \cite{bib:AaronsonArkhipov10}, known as boson-sampling, significantly simplifies the requirements for LOQC, allowing a type of non-universal quantum device that implements a classically hard algorithm using technologies that are, for the larger part, available today. In the boson-sampling model, we simply input $n$ copies of the single-photon Fock state into a passive linear optics network, comprised of beamsplitters and phase-shifters, and perform photodetection at the output. This yields a sampling problem, which is believed to be classically hard to simulate \cite{bib:AaronsonArkhipov10}. Thus, the full model requires only single-photon Fock state preparation, passive linear optics, and photodetection, technologies that are all available today on a small scale. For a review on various photon sources and photo-detectors see \cite{bib:SourceAndDetectorReview}. Note that number-resolving photo-detectors are not required in the limit of large $n$ as there is likely only zero or one photon per output mode \cite{bib:ark2012bos}. Several elementary experimental demonstrations have recently been performed \cite{bib:metcalf12, bib:Broome20122012, bib:Spring2, bib:Crespi3, bib:Tillmann4}.

In boson-sampling, the classical hardness of the sampling problem relates to the computational complexity of calculating the amplitudes in the output superposition. In the case of Fock state inputs, the output amplitudes are related to matrix permanents, the computation of which are known to be complete for the complexity class \textbf{\#P} and hence are not believed to be efficiently computable by classical means.

The classical hardness of this Fock state sampling raises the question as to whether there are other quantum states of light, which also yield classically hard sampling problems. It was shown recently by Seshadreesan \emph{et al.} \cite{bib:Kaushik14} that photon-added coherent states (PACS) and by Olson \emph{et al.} \cite{bib:olson2014sampling} that photon-added or photon-subtracted squeezed vacuum (PASSV) states are an example of such states. And it was shown by Lund et al. \cite{bib:lund2013boson} that a certain class of Gaussian state inputs yield a computationally hard sampling problem. It is known that passive linear optics may be efficiently simulated with Gaussian inputs and non-adaptive Guassian measurements \cite{bib:Bartlett02, bib:Bartlett02b}. However, the more general question as to which quantum states of light may be efficiently simulated with number-resolved measurements is an open question.

Given the result of AA for single photons, the natural question is whether this generalises to other quantum states. Here we consider a more generalized boson-sampling device where the input states are not Fock states, but rather superpositions of coherent states. This is a very broad class of continuous-variable optical states. 

We will structure this paper as follows. We first give a review of AA's boson-sampling formalism in terms of Fock and vacuum states and show the typical expression obtained at the output. Then, we analyse cat sampling in three separate limits:
\begin{enumerate}
\item First we analyse even and odd cat states and show that their Taylor expansions reduce to the vacuum and single-photon Fock state respectively as $\alpha\to0$. Thus, in the zero amplitude limit, cat sampling exactly reduces to boson-sampling and therefore yields a computationally hard problem. 
\item Second, we analyse small, but non-zero amplitude odd cat states. This is equivalent to Fock state sampling with some residual components that are treated as an error. This error is related to the AA proof for approximate boson-sampling, where it is required that the error rate satisfies a 1/poly(n) bound. Thus small, but non-zero, amplitude odd cat states are also computationally hard.
\item Third, we analyse general cat states which are arbitrary superpositions of two or more coherent states. We demonstrate that the output state is a highly entangled superposition of an exponential number of multi-mode coherent states \cite{bib:PhysRevA.46.2966, bib:PhysRevA.55.2478, bib:gerry2002nonlinear, bib:gerry2007nonlocal, bib:gerry2009maximally}, where the amplitude of each term is related to a permanent-like combinatoric problem, which would require exponential resources to compute via a brute-force approach. This provides strong evidence that such generalized optical sampling problems might in general be implementing classically hard problems. Determining a complete characterization of the computational complexity of such problems is a notoriously difficult open problem, but based on the evidence we present here, it likely resides in a classically hard class comparable to ideal boson-sampling.
\end{enumerate}

Next, to further support our evidence we present a complexity theoretic argument for the hardness of cat state sampling. We show that unless the polynomial hierarchy collapses to the third level there must not exist an efficient randomized classical algorithm which can produce an output distribution approximating that of an arbitrary interferometer with multiplicative error of $\sqrt{2}$ or less.

While such states may be more challenging to prepare than Fock states, addressing this question sheds light on what makes a quantum optical system classically hard to simulate, and may provide motivation for developing technologies for preparing quantum states of light beyond Fock states. We end this paper by discussing the prospects for experimentally preparing general cat states.

%
%

\section{Review of boson-sampling}

\begin{figure}[!htb]
\includegraphics[width=0.4\columnwidth]{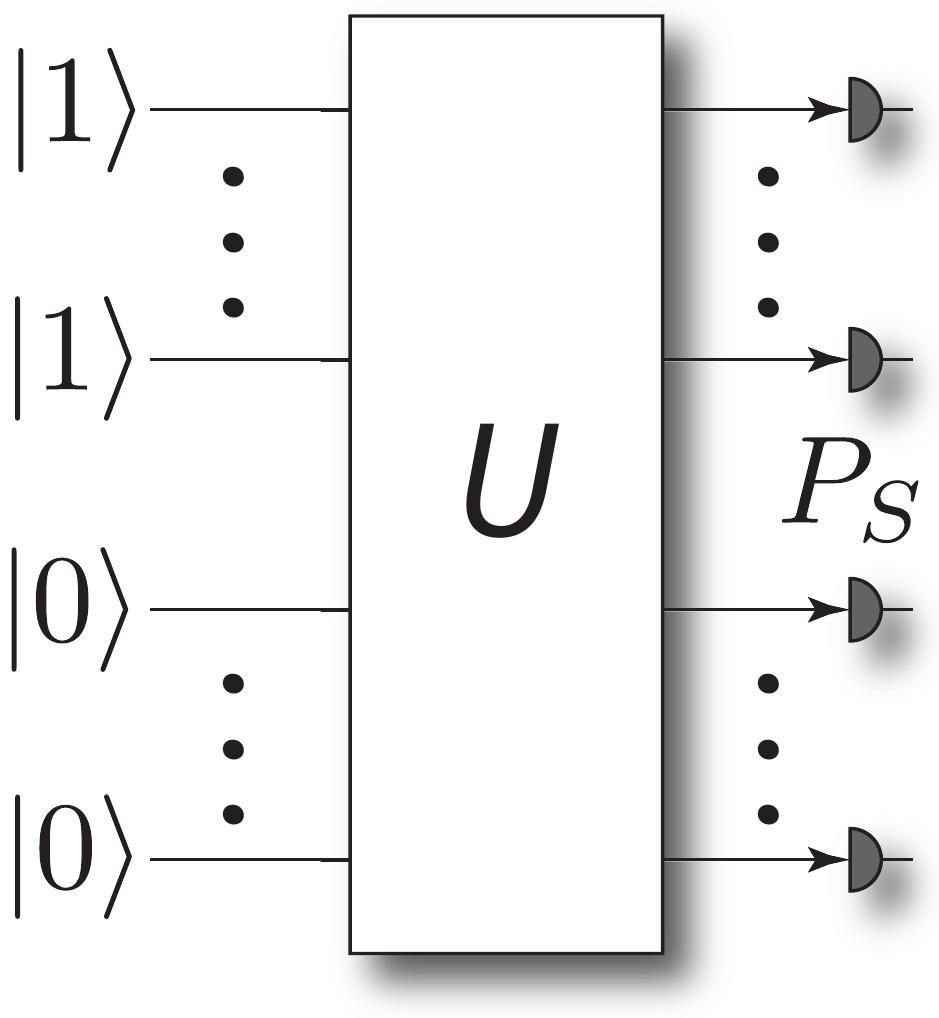}
\caption{The boson-sampling model for quantum computation. A series of single-photon and vacuum states are prepared, $\ket{1,\dots,1,0,\dots,0}$, and passed through a linear optics network, $\hat{U}$. The experiment is repeated many times and each time the output distribution is measured via coincidence number-resolved photodetection, sampling from the distribution $P_S$.} \label{fig:model}
\end{figure}

We begin by reviewing the boson-sampling model using single photons as illustrated in Fig.~\ref{fig:model}. For an elementary introduction to boson-sampling, see \cite{bib:GardBSintro}. First we prepare an $m$-mode state, where the first $n$ modes are prepared with the single photon Fock state, and the remainder with the vacuum state,
\begin{eqnarray} \label{eq:inputState}
\ket{\psi_\mathrm{in}} &=& \ket{1_1,\dots,1_n,0_{n+1},\dots,0_m} \nonumber \\
&=& a_1^\dag\dots a_n^\dag \ket{0_1,\dots,0_m},
\end{eqnarray}
where $a_i^\dag$ is the photon creation operator on the $i$th mode, and $m=O(n^2)$. Note that when $n$ photons are randomly inputed into the $m$ modes the sampling problem remains classically difficult \cite{bib:lund2013boson}. This has become colloquially known as scattershot boson-sampling. 

Next we propagate this state through a passive linear optics network, which can be expressed by a unitary map on the creation operators,
\begin{equation} \label{eq:unitary_map}
\hat{U}: \,\, \hat{a}_i^\dag \to \sum_{j=1}^m U_{i,j} \hat{a}_j^\dag.
\end{equation}

With this map, the output state can be expressed as,
\begin{equation}
\ket{\psi_\mathrm{out}} = \sum_S \gamma_S \ket{s_1,\dots, s_m},
\end{equation}
where $S$ are the different photon number configurations, $s_i$ is the number of photons in mode $i$ associated with configuration $S$, and \mbox{$\gamma_S\in \mathbb{C}$} are the respective amplitudes. The number of configurations scales exponentially with $n$, \mbox{$|S| = \binom{n+m-1}{n}$}. The total photon-number is conserved, thus \mbox{$\sum_i {s_i} = n$} for all $S$. Performing number-resolved photodetection, which are described by projection operators $\hat{\Pi}(n) = \ket{n}\bra{n}$, we sample from the distribution \mbox{$P_S=|\gamma_S|^2$}, each time obtaining an $m$-fold coincidence measurement outcome of a total of $n$ photons.

As shown by AA the sampling amplitudes are related to matrix permanents, 
\begin{eqnarray}
\gamma_S&=& \frac{\mathrm{Per}(U_S)}{\sqrt{\prod_{i=1}^m s_i! \prod_{j=1}^m t_j!}} \nonumber \\
&=& \frac{\sum_{\sigma\in S_n} \prod_{j=1}^n U_{j,\sigma_i}}{\sqrt{\prod_{i=1}^m s_i! \prod_{j=1}^m t_j!}},
\end{eqnarray}
where $U_S$ is an \mbox{$n \times n$} sub-matrix of $U$ as a function of the respective configuration $S$, $\mathrm{S}_n$ is the group of permutations, and $s_i$ ($t_j$) is the number of photons in the input (output) mode associated with mode $i$ ($j$). The best known classical algorithm for calculating matrix permanents is by Ryser \cite{bib:Ryser63}, requiring $O(2^n n^2)$ time steps. Because this requires exponential time to evaluate, sampling from the distribution $P_S$ is believed to be a classically hard problem. Importantly, boson-sampling does \emph{not} allow us to \emph{calculate} matrix permanents as this would require an exponential number of measurements.

%
%

\section{Zero Amplitude Cat analysis}

`Cat state' is a generic term for an arbitrary superposition of macroscopic states and may be used for quantum information processing \cite{bib:gilchrist2004schrodinger}. In quantum optics, this is generally understood to mean a superposition of two coherent states, potentially with large amplitudes. This is the definition we will use in this work. Two illustrative examples are the `even' ($+$) and `odd' ($-$) cat states, so-called because they contain only even or odd photon-number terms respectively,
\begin{equation}
\ket{\mathrm{cat}_\pm} = \frac{(\ket\alpha \pm \ket{-\alpha})}{\sqrt{2(1\pm e^{-2|\alpha|^2})}}.
\end{equation}


The odd cat state has the property that all of the even photon number terms vanish. In the limit of $\alpha\to0$ its amplitude identically approaches the single-photon state as shown here,
\begin{eqnarray}
\lim_{\alpha\to 0} \ket{\mathrm{cat}_-} &=& \lim_{\alpha\to0} \frac{\sqrt{2}e^{\frac{-|\alpha|^2}{2}}}{\sqrt{1-e^{-2|\alpha|^2}}}\left(\alpha\ket{1}+\frac{\alpha^3\ket{3}}{\sqrt{3!}}+\dots\right) \nonumber \\
&\approx& \ket{1}+O(\alpha^2)\ket{3} \nonumber \\
&\to& \ket{1}.
\end{eqnarray}
In the limit as $\alpha\to0$ we ignore all higher order $\alpha$ terms. 

Furthermore, the vacuum state (we require $O(n^2)$ vacuum states to be consistent with the boson-sampling model) is given by a trivial cat state containing only a single term in the superposition (\mbox{$t=1$}) with the respective amplitude \mbox{$\alpha=0$}. Alternately, the vacuum state can be regarded as the zero amplitude limit of the even cat state,
\begin{eqnarray}
\lim_{\alpha\to 0} \ket{\mathrm{cat}_+} &=& \lim_{\alpha\to0} \frac{\sqrt{2}e^{\frac{-|\alpha|^2}{2}}}{\sqrt{1+e^{-2|\alpha|^2}}}\left(\alpha\ket{0}+\frac{\alpha^2\ket{2}}{\sqrt{2!}}+\dots\right) \nonumber \\
&\approx& \ket{0}+O(\alpha^2)\ket{2} \nonumber \\
&\to& \ket{0}.
\end{eqnarray}

Thus, it is immediately clear that in the $\alpha\to0$ amplitude limit, cat state sampling reduces to ideal boson-sampling, using an appropriate configuration of odd and even cat states, which is a provably hard problem. We use the term `provably hard' to mean computationally hard, assuming that ideal and approximate boson-sampling are computationally hard. Specifically, to implement exact boson-sampling with cat states, we choose our input state to be,
\begin{eqnarray}
\ket{\psi_\mathrm{in}} &=& \lim_{\alpha\to 0} (\ket{\mathrm{cat}_-}_1\dots \ket{\mathrm{cat}_-}_n \ket{\mathrm{cat}_+}_{n+1}\dots \ket{\mathrm{cat}_+}_m) \nonumber \\
&=& \ket{1_1,\dots,1_n,0_{n+1},\dots,0_m},
\end{eqnarray}
which is exactly the form of Eq. \ref{eq:inputState}. This example is trivial but the point is to show a simple example of cat states leading to a computationally hard problem in a particular limit, which raises the question as to whether it remains hard as we transition out of that limit. In App.~\ref{app:odd_cat} we present an example of this reduction in the case of Hong-Ou-Mandel interference to explicitly demonstrate that small amplitude cats behave as single photons. This demonstrates that in certain regimes, cat state sampling reproduces single-photon statistics.

\section{small amplitude cat analysis}
Having established that cat sampling reduces to boson-sampling in the zero amplitude limit, the obvious next question is `what if the amplitude is small but non-zero?'. It was shown by AA that boson-sampling, when corrupted by erroneous samples, remains computationally hard provided that the error rate scales as $1/\mathrm{poly}(n)$. If we consider a small, but non-zero, amplitude odd cat state, we can treat the non-single-photon terms, which scale as a function of $\alpha$, as erroneous terms. The error that these erroneous terms induce must be kept below the $1/\mathrm{poly}(n)$ bound. Specifically,
\begin{equation}
\ket{\mathrm{cat}_-} = \underbrace{\gamma_1(\alpha)\ket{1}}_{\mathrm{single\,photon}} + \underbrace{\gamma_3(\alpha) \ket{3} + \dots}_{\mathrm{error\,terms}},
\end{equation}
where $\gamma_i(\alpha)$ defines the odd photon-number distribution and follows from Eq.~\ref{eq:fna}. The two underbraced components represent the desired single-photon term and the remaining photon-number terms, which are treated as errors. 

In App.~\ref{app:poly_bound} we show that the bound on the amplitude of the cat states for a provably hard sampling problem to take place is,
\begin{equation} \label{eq:bound_alpha_hard}
\alpha^{2n}\mathrm{csch}^n(\alpha^2) > 1/\mathrm{poly}(n),
\end{equation}
where we input odd cat states in every mode requiring a $\ket{1}$ and vacuum in the remaining modes. Although this function is exponential in $n$ the probability of successfully sampling from the correct distribution will satisfy this bound for sufficiently small values of $n$ and $\alpha$. The value of $n$ may still be large enough however to implement a post-classical boson-sampling device. Thus, it follows that for non-zero, but sufficiently small $\alpha$, cat sampling remains computationally hard.

We have established that cat state boson-sampling is a provably computationally hard problem in two regimes: (1) in the $\alpha\to0$ amplitude limit, in which case we reproduce ideal boson-sampling, and (2) for non-zero but sufficiently small amplitudes, in which case the non-single-photon-number terms may be regarded as errors, which remains a computationally hard problem, subject to the bound given in Eq.~\ref{eq:bound_alpha_hard}. Having established this, the remainder of this paper is dedicated to the completely general case, whereby the terms in the cat states may have arbitrary amplitude, potentially at a macroscopic scale.

%
%

\section{Arbitrary amplitude cat analysis}

In this section we will consider arbitrary superpositions of an arbitrary number of coherent states, in which case a general cat is of the form,
\begin{equation}
\ket{\mathrm{cat}} = \sum_{j=1}^t \lambda_j \ket{\alpha_j}.
\end{equation}

Let the input state to our generalized boson-sampling model comprise $m$ arbitrary superpositions of $t$ coherent states, which we will refer to as generalized cat states,
\begin{equation}
\ket{\psi_\mathrm{in}} = \bigotimes_{i=1}^m \sum_{j=1}^t \lambda_j^{(i)}\ket{\alpha_{j}^{(i)}},
\end{equation}
where $\ket{\alpha_{j}^{(i)}}$ is the coherent state of amplitude \mbox{$\alpha\in\mathbb{C}$} of the $j$th term in the $i$th mode, and \mbox{$\lambda_j^{(i)}\in\mathbb{C}$} is the amplitude of the $j$th term of the superposition in the $i$th mode\footnote{Continuous superpositions are a simple generalization of our formalism, and with this generalization arbitrary states could be expressed as continuous superpositions of coherent states.}. It should be noted here that, in line with traditional boson-sampling, we can choose a number of the modes to be the vacuum. This is achieved by setting \mbox{$\lambda_1^{(i)}=1$} and \mbox{$\alpha_1^{(i)}=0$}.

\begin{figure}[!htb]
\includegraphics[width=0.65\columnwidth]{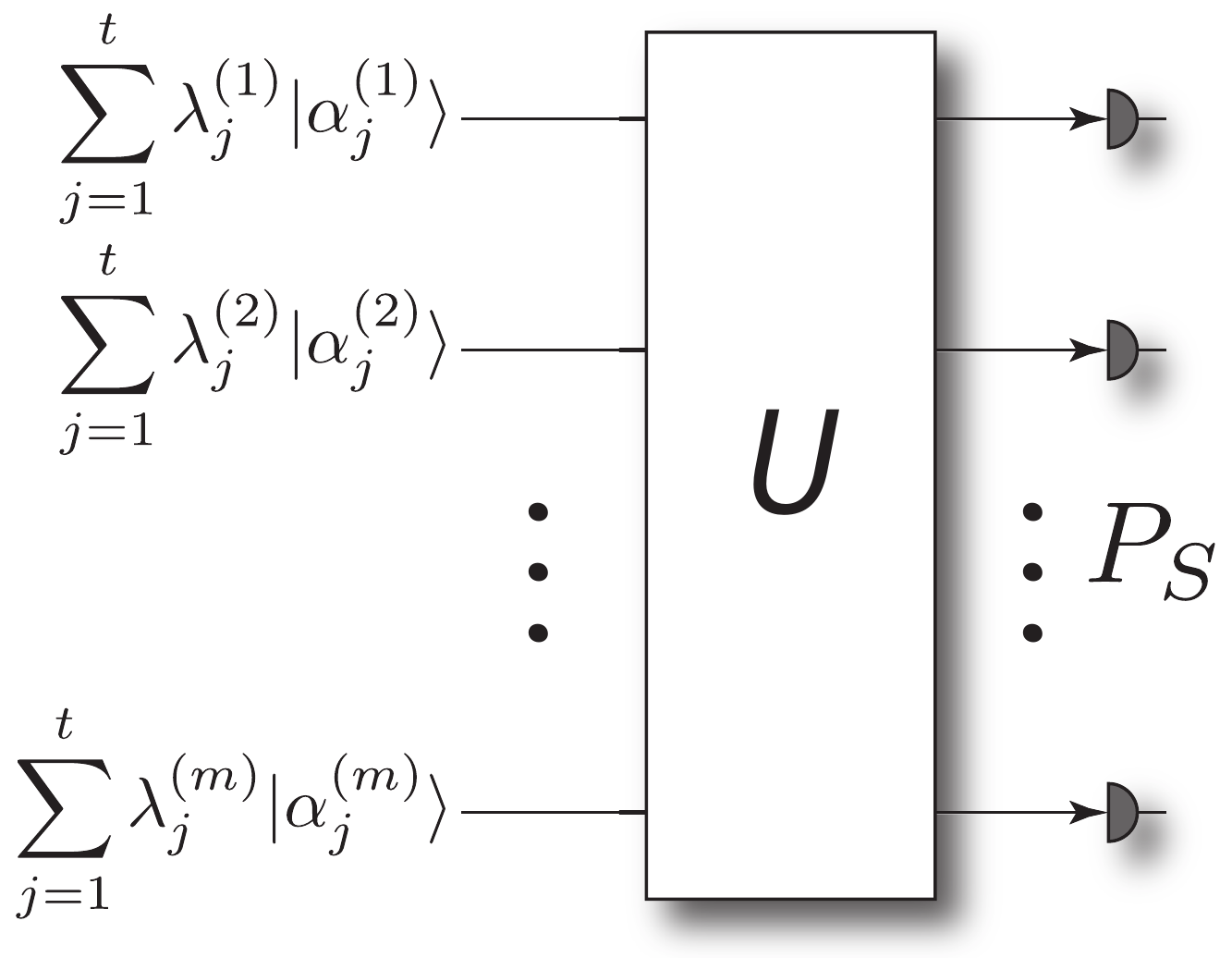}
\caption{The model for generalized boson-sampling with superpositions of coherent states --- `cat states'. The input state to each mode is an arbitrary superposition of coherent states, some of which are set to the vacuum. Following the application of a linear optics network, the distribution is sampled via number-resolved photo-detection.} \label{fig:cat_model}
\end{figure}

Expanding this expression yields a superposition of multi-mode coherent states of the form,
\begin{equation} \label{eq:psi_in}
\ket{\psi_\mathrm{in}} = \sum_{\vec{t}=1}^{t} \lambda_{t_1}^{(1)}\dots \lambda_{t_m}^{(m)} \ket{\alpha_{t_{1}}^{(1)}, \dots, \alpha_{t_{m}}^{(m)}},
\end{equation}
where $\vec{t}$ is shorthand for $\{t_1,...,t_m\}$. We propagate this state through the passive linear optics network $\hat{U}$ illustrated in Fig.~\ref{fig:cat_model}. Such a unitary network has the property that a multi-mode coherent state is mapped to another multi-mode coherent state,
\begin{equation} \label{eq:coherent_map}
\hat{U} \ket{\alpha^{(1)},\dots,\alpha^{(m)}} \to \ket{\beta^{(1)},\dots,\beta^{(m)}},
\end{equation}
where the relationship between the input and output amplitudes is given by (see App. \ref{app:coherent_map}),
\begin{equation} \label{eq:coherent_map_relation}
\beta^{(j)} = \sum_{k=1}^m U_{j,k} \alpha^{(k)}.
\end{equation}
$\hat{U}$ acts on each term in the superposition of Eq.~\ref{eq:psi_in} independently. Thus, the output state will be of the form,
\begin{eqnarray} \label{eq:psi_out}
\ket{\psi_\mathrm{out}} &=& \hat{U} \ket{\psi_\mathrm{in}} \nonumber \\
&=& \sum_{\vec{t}=1}^{t} \lambda_{t_1}^{(1)}\dots \lambda_{t_m}^{(m)} \ket{\beta_{\vec{t}}^{(1)}, \dots, \beta_{\vec{t}}^{(m)}}.
\end{eqnarray}
The number of terms in the output superposition is $t^m$, scaling exponentially with the number of modes, provided \mbox{$t>1$}.

Our goal is to sample this distribution using number-resolved photodetectors, which are described by the measurement projectors,
\begin{equation}
\hat\Pi_i(n) = \ket{n}_i\bra{n}_i,
\end{equation}
where $n$ is the photon-number measurement outcome on the $i$th mode. Multi-mode measurements are described by the projectors,
\begin{equation} \label{eq:projector_S}
\hat\Pi(S) = \hat\Pi_1(S_1) \otimes \dots \otimes \hat\Pi_m(S_m),
\end{equation}
where \mbox{$S=\{S_1,\dots,S_m\}$} is the multi-mode measurement signature, with $S_i$ photons measured in the $i$th mode. The sample probabilities are given by,
\begin{equation} \label{eq:sample_prob}
P_S = \bra{\psi_\mathrm{out}} \hat\Pi(S) \ket{\psi_\mathrm{out}}.
\end{equation}
In the case of continuous-variable states, the number of measurement signatures, $|S|$, is unbounded as the photon-number is undefined, unlike Fock states where the total photon-number is conserved. 

Without presenting a rigorous complexity argument, we argue that this sampling problem is likely to be classically hard if three intuitive criteria are satisfied:
\begin{enumerate}
\item There must be an exponential number of terms in the output distribution. This rules out brute-force simulation by explicitly calculating the state vector.
\item The terms in the superposition must be entangled, such that the distribution cannot be trivially sampled by independently sampling each mode.
\item Each of the amplitudes in the output distribution must be related to a computationally hard problem. This ensures that classical simulation of the individual amplitudes is not efficient.
\end{enumerate}

We have chosen these criteria as they are general properties that classically hard problems are known to exhibit, but we do not prove that these criteria are sufficient to establish whether a problem is likely to be classically hard. For example, ideal boson-sampling is known to be computationally hard, satisfying all 3 criteria. However, fermionic-sampling is known to be classically efficient, and violates criteria 3, as it relates to matrix determinants rather than permanents, which reside in \textbf{P}.

Criteria (1) is achieved by virtue of our choice of input state --- there are $t^m$ terms in the output distribution.

It is easily seen that criteria (2) holds in general. As a simple example, consider the input state,
\begin{equation}
\ket{\psi_\mathrm{in}} = \mathcal{N}^2 (\ket\alpha + \ket{-\alpha})\otimes (\ket\alpha + \ket{-\alpha}) = \ket{\mathrm{cat_+},\mathrm{cat_+}},
\end{equation}
a tensor product of two even cat states. Passing this separable two-mode state through a 50/50 beamsplitter gives rise to the output state,
\begin{eqnarray}
\ket{\psi_\mathrm{out}} = \hat{H}\ket{\psi_\mathrm{in}} = \ket{\mathrm{cat}',0} + \ket{0,\mathrm{cat}'},
\end{eqnarray}
where \mbox{$\ket{\mathrm{cat}'}=\mathcal{N}^{2}_{+}(\ket{\sqrt{2}\alpha} + \ket{-\sqrt{2}\alpha})$} is a cat state. This is a path-entangled superposition of a cat state across two modes. Thus, while Eq.~\ref{eq:coherent_map} demonstrates that a unitary network maps a tensor product of coherent states to a tensor product of coherent states, such a network will in general generate path-entanglement when the input state is a tensor product of superpositions of coherent states. Note the structural similarity between cat state interference and two-photon Hong-Ou-Mandel (HOM) \cite{bib:HOM87} -type interference. In the case of HOM interference we have \mbox{$\hat{H}\ket{1,1}= (\ket{2,0}+\ket{0,2})/\sqrt{2}$}, whereas for cat states we have \mbox{$\hat{H}\ket{\mathrm{cat},\mathrm{cat}}= \ket{\mathrm{cat}',0} + \ket{0,\mathrm{cat}'}$}.

It was recently and independently reported by Jiang \emph{et al.} \cite{bib:Jiang13} that linear optics networks fed with nonclassical pure states of light almost always generates modal entanglement, consistent with our observation here. This ensures that the output state to our generalized boson-sampling device is highly entangled, thus satisfying criteria (2). However, Jiang \textit{et al.} present no discussion about our hardness criteria (3); they do not connect their states to a computationally hard problem. Thus their work provides a necessary but not sufficient proof of computational hardness. It is important, as in our work here, to examine such non-classical input states individually and make the case for the importance of criteria (3). For example it is well known from the Gottesman-Knill theorem that some systems with exponentially large Hilbert spaces that satisfy our criteria (1) and (2) can nevertheless be efficiently simulated. An example is the circuit model for quantum computation that deploys only gates from the Clifford algebra.

Finally let us consider criteria (3). Let the expansion for a coherent state be,
\begin{equation}
\ket\alpha = \sum_{n=0}^\infty f_n(\alpha)\ket{n},
\end{equation}
in the photon-number basis, where,
\begin{equation} \label{eq:fna}
f_n(\alpha) = e^{-\frac{|\alpha|^2}{2}} \frac{\alpha^n}{\sqrt{n!}},
\end{equation}
is the amplitude of the $n$-photon term. Then,
\begin{equation}
\langle n|\alpha\rangle = f_n(\alpha).
\end{equation}
Thus, acting the measurement projector for configuration $S$, Eq.~\ref{eq:projector_S}, on the output state, Eq.~\ref{eq:psi_out}, we obtain, 
\begin{equation}
\hat\Pi(S) \ket{\psi_\mathrm{out}} = \gamma_S \ket{S_1, S_2, \dots,S_m},
\end{equation}
where,
\begin{eqnarray} \label{eq:gamma}
\gamma_S = \sum_{\vec{t}=1}^{t} \left(\prod_{j=1}^m \lambda_{t_j}^{(j)} f_{S_j}\!\!\left(\sum_{k=1}^m U_{j,k} \alpha_{t_j}^{(k)}\right) \right),
\end{eqnarray}
and the sampling probability takes the form \mbox{$P_S = |\gamma_S|^2$}. We can group the terms under the product and label them $A_{j,\vec{t}}^{(S)}$. Then the amplitudes are given by,
\begin{equation} \label{eq:S_perm}
\gamma_S = \sum_{\vec{t}=1}^t \prod_{j=1}^m A_{j,\vec{t}}^{(S)},
\end{equation}
which has the same analytic structure as the permanent when $t=m$ but sums over additional terms that are not present in the permanent. Evaluating this combinatoric problem requires exponential resources using brute-force. Via brute force, evaluating this expression requires summing $t^m$ terms. Given that Eq. \ref{eq:S_perm} has the same analytic form as the matrix permanent, which is known to be classically hard, this implies a striking similarity between cat state sampling and Fock state sampling, with the constraint that $A$ is of a form whose permanent is not trivial. In fact, in the $\alpha\to0$ limit, evaluating this combinatoric expression \emph{must} be as hard as calculating an $n\times n$ matrix permanent, since we know that in this limit the problem reduces to ideal boson-sampling. In the original proof by Aaronson \& Arkhipov, it is required that $U$ is Haar-random. It is an open question as to whether $A$ can be made Haar-random in the presented generalized boson-sampling model.

In the trivial case of \mbox{$t=1$} this expression simplifies to,
\begin{equation}
\gamma_S = \prod_{j=1}^m f_{S_j}\!\!\left(\sum_{k=1}^m U_{j,k} \alpha_{1}^{(k)}\right),
\end{equation}
which evaluates in polynomial time. In this case the input state is simply a tensor product of coherent states, and the runtime is consistent with the known result that simulating coherent states is trivial as the tensor product structure allows sampling to proceed by independently sampling each mode, each of which is an efficient sampling problem. However, when \mbox{$t>1$} the complexity of directly arithmetically evaluating Eq.~\ref{eq:gamma} grows exponentially.

%
%

\section{A computational complexity argument} \label{sec:comp_sci_arg}

We now provide a complexity theoretic argument for the hardness of cat-state sampling. We will consider the problem of weakly simulating such a system with a randomized classical algorithm, and show that the existence of any such algorithm which can approximate the output to within a constant multiplicative factor of $\sqrt{2}$ would imply a collapse of certain classical computational complexity classes which are believed to be distinct. The polynomial hierarchy, denoted as \textbf{PH}, is composed of an infinite number of levels $k$, which are composed of the complexity classes \mbox{$\Sigma_k \mathbf{P} = \mathbf{NP}^{\Sigma_{k-1} \mathbf{P}}$}, \mbox{$\Delta_k \mathbf{P} = \mathbf{P}^{\Sigma_{k-1} \mathbf{P}}$}, and \mbox{$\Pi_k \mathbf{P} = \mathbf{coNP}^{\Sigma_{k-1} \mathbf{P}}$}, where \mbox{$\Sigma_0 \mathbf{P} = \Delta_0 \mathbf{P} = \Pi_0 \mathbf{P} = \mathbf{P}$}. We show that if the measurement results in a cat state interferometry experiment are specified by a string $x$, and each $x$ occurs with probability $p(x)$, then if there exists an efficient (polynomial time) randomized algorithm which, for all such cat state interferometry experiments, produces output $x$ with probability $p'(x)$ such that \mbox{$\frac{1}{\sqrt{2}} p(x) \leq p'(x) \leq \sqrt{2} p(x)$} for all $x$, then the polynomial hierarchy collapses to the third level (i.e. \mbox{$\mathbf{PH} = \Delta_3 \mathbf{P}$}). While such a result would not be quite as strong as \mbox{$\mathbf{P}=\mathbf{NP}$}, it has a well established connection to this problem, and it is widely believed that the polynomial hierarchy has an infinite number of inequivalent levels. 

Our approach follows the method developed by Bremner, Jozsa \& Shepherd \cite{bib:bremner2011classical}, which they used to show the intractability of classical simulation of certain circuits composed of commuting gates. The technique has since been extended to other settings, including the one clean qubit (DQC1) model \cite{bib:morimae2014hardness,bib:morimae2014classical}. This approach makes use of the notion of post-selection, in which the computational cost of performing a certain computation is counted only if that computation results in a particular value of a chosen post-selection register.

We provide informal definitions of two computational complexity classes based on post-selection, based on those used in \cite{bib:morimae2014hardness}, which will be used extensively in our proof. More formal definitions of these classes can be found in \cite{bib:bremner2011classical}. These classes are defined in terms of circuits composed of either classical or quantum gates with identified output and post-selection registers $O_x$ and $P_x$ respectively. A language $L$ is in the class $\mathbf{PostBPP}$ if and only if there exists a uniform family of classical circuits and a \mbox{$0<\delta<1/2$} such that:
\begin{enumerate}
\item if $x \in L$ then $\text{Prob}(O_x=1|P_x = 0\dots 0) \geq \frac{1}{2}+\delta$,
\item if $x \notin L$ then $\text{Prob}(O_x=1|P_x = 0\dots 0) \leq \frac{1}{2}-\delta$.
\end{enumerate}
Similarly, a language $L$ is in the class \textbf{PostBQP} if there exist a uniform family of quantum circuits satisfying the above criteria \footnote{Here the gate set is assumed to be some standard universal set, such as the Hadamard and Toffoli gates. More exotic gate sets can potentially lead to more powerful models, but this class provides a lower bound on the power of post-selected quantum computation, which is sufficient for our purposes.}. We also define a third complexity class \textbf{PostCAT} to capture the notion of post-selection applied to the interferometry experiments we are concerned with. We will say that a language $L$ is in \textbf{PostCAT} if there exists a uniform family of $n$-port linear interferometers acting on cat state inputs satisfying the preceding criteria.

The complexity of the first two classes has previously been studied, and it is known that \textbf{PostBPP} exactly corresponds to another complexity class known as $\mathbf{BPP}_\text{path}$, which is contained within \mbox{$\Delta_3 \mathbf{P}$} \cite{bib:han1997threshold}. On the other hand, it was shown by Aaronson that \mbox{$\mathbf{PostBQP} = \mathbf{PP}$} \cite{bib:aaronson2005quantum}. This is rather surprising, since \mbox{$\mathbf{PH} \subseteq \mathbf{P}^\mathbf{PP}$}, which implies a difference between the power of post-selected classical and quantum computation unless \mbox{$\mathbf{P}^\mathbf{PP} = \mathbf{PH} = \Delta_3 \mathbf{P}$}. Following from the definitions of \textbf{PostBPP} and \textbf{PostBQP} given above, this further implies that an efficient classical randomized algorithm cannot mimic the output of an arbitrary quantum circuit, since such a situation would yield \mbox{$\mathbf{PostBQP}=\mathbf{PostBPP}$} and so collapse the polynomial hierarchy. In fact, this last statement can be made stronger: The existence of a polynomial time randomized classical algorithm which approximates the output distribution to an arbitrary quantum circuit to within multiplicative error of $\sqrt{2}$ would yield \mbox{$\mathbf{PH} = \Delta_3 \mathbf{P}$} \cite{bib:bremner2011classical,bib:morimae2014hardness}. 

All that must be done, then, in order to prove our claim, is to show that $\mathbf{PostBQP} \subseteq \mathbf{PostCAT}$, from which it would follow that the existence of an efficient randomized classical algorithm which can approximate the output distribution of an arbitrary linear interferometer applied to cat state inputs to within a multiplicative error of $\sqrt{2}$ would imply that \mbox{$\mathbf{PostCAT} \subseteq \mathbf{PostBPP}$} and hence \mbox{$\mathbf{PH} = \Delta_3 \mathbf{P}$}. However, that \mbox{$\mathbf{PostBQP} \subseteq \mathbf{PostCAT}$} follows directly from the work of Ralph \emph{et al.} \cite{bib:ralph}, who showed that arbitrary quantum circuits could be implemented exactly, albeit probabilistically, on qubits encoded as a superposition of even and odd parity cat states. Although the set of gates they introduce is probabilistic, there is always some probability of obtaining a measurement result which correctly implements the desired gate with unit fidelity. Hence by post-selecting on such an outcome, the system can be made to deterministically implement an arbitrary quantum circuit. Since post-selection could also be applied to output qubits from the circuit while remaining in \textbf{PostCAT}, it follows that any computation in \textbf{PostBQP} is also in \textbf{PostCAT}. Thus we must conclude that unless \mbox{$\mathbf{PH} = \Delta_3 \mathbf{P}$}, there must not exist an efficient randomized classical algorithm which can produce an output distribution approximating that of an arbitrary interferometer with multiplicative error of $\sqrt{2}$ or less.

%
%

\section{Preparing cat states}

Finally, we will discuss the prospects for experimentally preparing cat states of the form used in our derivation. There exists a significant number of schemes for generating a finite number of superpositions of coherent states all of which are extremely difficult to scale to higher order cat states. For example, superpositions of coherent states with equal amplitudes but different phases can be produced with quantum nondemolition (QND) measurements \cite{bib:PhysRevA.37.2970} via the interaction of a strong Kerr nonlinearity \cite{bib:Haroche06, bib:GerryKnight05}. Another approach is to use strong Kerr nonlinearities together with coupled Mach-Zehnder interferometers \cite{bib:GerryKnight05} but this is impractical as outside the cavity a strong Kerr would require a coherent Electromagnetically Induced Transparency \cite{bib:harris2008, bib:PhysRevLett.64.1107} effect in an atomic gas cloud and even there in practice the nonlinearities are too weak for our purposes.

In a similar way that measurements of photon number can produce discrete coherent state superpositions in phase; measurements of the phase can produce discrete coherent state superpositions in amplitude. This can be understood via the number-phase uncertainty relation. Any improved knowledge of the phase of a state induces kicks in the number and vice versa. In this way, by combining such different measurements, one can produce discrete superpositions in both phase and amplitude; approaching the arbitrary superpositions of coherent states we require here. Exactly such a scheme was proposed by Jeong \emph{et al.} in 2005 \cite{bib:Lund04, bib:JeongLundRalph05}. By combining both types of detection schemes, even with detectors of non-unit efficiency, they show that a large number of propagating superpositions of coherent states may be thus produced. These states then could be used in proof-of-principle experiments for our protocol outlined here.
%
%

\section{Conclusion}

We have presented evidence that a linear optics network, fed with arbitrary superpositions of coherent states, and sampled via number-resolved photodetection, is likely to be a classically hard problem. We have shown that sampling within multiplicative errors is hard and we have also presented evidence that it has similar complexity to boson-sampling, a model for which sampling with even additive error is known to be hard. Our argument is based on three realistic criteria for computational hardness of the sampling problem. In the case of input states comprising superpositions of coherent states, these three criteria are satisfied. In the case of criteria (3), we find that the amplitudes are related to a permanent-like function of a matrix, strikingly similar to ideal boson-sampling. In fact, if this permanent-like function is shown to be in the Haar random class of matrices, then our result is provably hard following Aaronson and Arkhipov's proof.

Furthermore, we show two examples of how sampling with normal cat states reduces to a computationally hard problem,
\begin{enumerate}
\item In the limit of $\alpha\to0$ amplitude cat states, cat sampling reduces to ideal boson-sampling.
\item When the amplitude is increased slightly away from zero, odd cat state sampling is hard for sufficiently small amplitudes following the bound of Eq. \ref{eq:bound_alpha_hard}. We show this by treating the non-single-photon-number terms as an error model.
\end{enumerate}
Furthermore, given that (1) is known to be computationally hard, as it reduces to ideal boson-sampling, it follows that our combinatoric expression for evaluating the amplitudes in the output superposition, Eq. \ref{eq:S_perm}, is equivalent to evaluating a permanent in this limit. For arbitrary alpha, the combinatoric expression maintains the same analytic form. This presents strong evidence that arbitrarily large cat states yield a hard sampling problem akin to standard boson-sampling. 

With all of these results combined, these observations present strong evidence that such a generalized sampling problem is likely classically hard to simulate and is indeed hard within multiplicative errors.

Because coherent states form an over-complete basis, any pure optical state can be expressed in terms of coherent states, suggesting that most quantum states of light may yield hard sampling problems. This observation further motivates interest in developing sources for quantum states of light other than Fock states. \\

%
%

\begin{acknowledgments}
We would like to thank Mark Wilde for helpful discussions. This research was conducted by the Australian Research Council Centre of Excellence for Engineered Quantum Systems (Project number CE110001013). JPD would like to acknowledge the Air Force Office of Scientific Research, the Army Research Office, and the National Science Foundation. This work was partly supported by DSTL (contract number DSTLX1000063869). This material is based in part on research supported in part by the Singapore National Research Foundation under NRF Award No. NRF-NRFF2013-01.
\end{acknowledgments}

%
%

\bibliography{bibliography}

%
%

\appendix

%
%

\section{Propagating multi-mode coherent states through passive linear optics networks} \label{app:coherent_map}

The unitary map describing a passive linear optics network is given by,
\begin{equation} \label{eq:app_unitary_map}
\hat{U}: \,\, \hat{a}_i^\dag \to \sum_{j=1}^m U_{i,j} \hat{a}_j^\dag,
\end{equation}
and taking the Hermitian conjugate yields,
\begin{equation}
\label{eq:app_unitary_map_conj}
\hat{U}: \,\, \hat{a}_i \to \sum_{j=1}^m U_{i,j}^* \hat{a}_j.
\end{equation}
A coherent state can be expressed in terms of a displacement operator acting on the vacuum state,
\begin{equation}
\ket{\alpha^{(i)}}_i = \hat{D}_i(\alpha^{(i)}) \ket{0}_i,
\end{equation}
where the displacement operator may be expressed in terms of creation and annihilation operators as,
\begin{equation}
\hat{D}_i(\alpha^{(i)}) = \mathrm{exp}(\alpha^{(i)} \hat{a}_i^\dag - {\alpha^{(i)}}^* \hat{a}_i).
\end{equation}
Applying the unitary map Eqs.~\ref{eq:app_unitary_map} \& \ref{eq:app_unitary_map_conj}, we obtain,
\begin{equation}
\hat{U}\hat{D}_i(\alpha_i) = \mathrm{exp}\!\left(\alpha^{(i)} \sum_{j=1}^m U_{i,j} \hat{a}_j^\dag - {\alpha^{(i)}}^* \sum_{j=1}^m U_{i,j}^* \hat{a}_j\right).
\end{equation}
Let,
\begin{equation}
\hat{U} \ket{\alpha^{(1)},\dots,\alpha^{(m)}} = \ket{\beta^{(1)},\dots,\beta^{(m)}}.
\end{equation}
Then,
\begin{equation}
\hat{U} \ket{\alpha^{(1)},\dots,\alpha^{(m)}} = \hat{U} \hat{D}_1(\alpha^{(1)}) \dots \hat{D}_m(\alpha^{(m)}) \ket{0_1,\dots,0_m}.
\end{equation}
For each term,
\begin{eqnarray}
\hat{U} \hat{D}_i(\alpha^{(i)}) &=& \prod_{j=1}^m \mathrm{exp}\!\left(\alpha^{(i)} U_{i,j} \hat{a}_j^\dag - {\alpha^{(i)}}^* U_{i,j}^* \hat{a}_j\right) \nonumber \\
&=& \prod_{j=1}^m \hat{D}_j(U_{i,j} \alpha^{(i)}).
\end{eqnarray}
Thus,
\begin{eqnarray}
\hat{U} \prod_{i=1}^m \ket{\alpha^{(i)}}_i &=& \hat{U} \prod_{i=1}^m \hat{D}(\alpha^{(i)}) \ket{0}_i \nonumber \\
&=& \prod_{i=1}^m \prod_{j=1}^m \hat{D}_j(U_{i,j} \alpha^{(i)}) \ket{0} \nonumber \\
&=& \bigotimes_{j=1}^m \left|\sum_{i=1}^m U_{i,j} \alpha^{(i)}\right\rangle_j \nonumber \\
&=& \bigotimes_{j=1}^m \ket{\beta^{(j)}}_j.
\end{eqnarray}
And,
\begin{equation}
\label{eq:final_beta}
\beta^{(j)} = \sum_{i=1}^m U_{i,j} \alpha^{(i)},
\end{equation}
as per Eq.~\ref{eq:coherent_map_relation}. 

%
%

\section{Reproducing Hong-Ou-Mandel interference using small amplitude odd cat states} \label{app:odd_cat}

We begin with our generalized cat state result from Eq. \ref{eq:gamma}. 
\begin{equation} \label{eq:mainresult}
\gamma_s = \sum_{\vec{t}=1}^{t} \left(\prod_{j=1}^m \lambda_{t_j}^{(j)} f_{S_j}(\beta_{\vec{t}}^{(j)})\right),
\end{equation}
and input the odd cat state which has the form 
\begin{eqnarray} \label{eq:app_odd_cate_state}
\ket{\mathrm{cat}_-} &=& \frac{\ket{\alpha}-\ket{-\alpha}}{\sqrt{2(1-\mathrm{exp}[-2\alpha^2])}}.
\end{eqnarray}

When considering the specific example of $\ket{\mathrm{cat}_-}$ the $\lambda_{t_j}^{(j)}$ of Eq.~\ref{eq:mainresult} goes to $(-1)^{t_j}$. Eq.~\ref{eq:mainresult} then becomes,
\begin{eqnarray} \label{eq:mainresult2}
\gamma_s &=& \sum_{\vec{t}=1}^{t} \left(\prod_{j=1}^m (-1)^{t_j} \frac{f_{S_j}(\beta_{\vec{t}}^{(j)})}{\sqrt{2(1-\mathrm{exp}[-2\alpha^2])}} \right).
\end{eqnarray}
Since the $\beta_{\vec{t}}^{(j)}$'s in Eq.~\ref{eq:mainresult2} depend on $\alpha$, we substitute the argument of $f_{S_j}$ using Eq.~ \ref{eq:fna},
\begin{eqnarray} 
\gamma_s &=& \frac{1}{\left(\sqrt{2(1-\mathrm{exp}[-2\alpha^2])}\right)^m} \\
&\times& \sum_{\vec{t}=1}^{t} \left(\prod_{j=1}^m (-1)^{t_j}  \mathrm{exp}\left[-\frac{|\beta_{\vec{t}}^{(j)}|^2}{2}\right] \frac{(\beta_{\vec{t}}^{(j)})^{S_j}}{\sqrt{S_j!}}\right)
\end{eqnarray}
Next we take a first order approximation. Since $\alpha$ is small, the exponential in the numerator goes to one while the exponential in the denominator goes to $\mathrm{exp}(x)\approx 1+x$ because otherwise this would diverge. This yields,
\begin{eqnarray} \label{eq:sub}
\gamma_s &\approx& \frac{1}{\left(\sqrt{2(1-(1-2\alpha^2)}\right)^m} \sum_{\vec{t}=1}^{t} \left(\prod_{j=1}^m (-1)^{t_j} \left(1\right) \frac{(\beta_{\vec{t}}^{(j)})^{S_j}}{\sqrt{S_j!}}\right) \nonumber \\
&=& \frac{1}{(2\alpha)^m\sqrt{S_1!S_2!\dots S_m!}} \sum_{\vec{t}=1}^{t} \left(\prod_{j=1}^m (-1)^{t_j} (\beta_{\vec{t}}^{(j)})^{S_j}\right) \nonumber \\
&=& \frac{1}{(2\alpha)^m\sqrt{S_1!S_2!\dots S_m!}} \sum_{\vec{t}=1}^{t} (-1)^{\sigma(\vec{t})}\prod_{j=1}^m (\beta_{\vec{t}}^{(j)})^{S_j}.
\end{eqnarray}

In the limit of small $\alpha$ we know that the odd cat state reduces to a single photon Fock state. Here we consider the case of a cat state being inputted into the first two modes and let the unitary be the Hadamard gate. In small $\alpha$ this corresponds to inputting a single photon Fock state into the first two modes and interfering them in a single 50/50 beamsplitter. Therefore, the corresponding bunching in the output modes would to be expected. In this section we show that our expression of Eq.~\ref{eq:sub} does show the expected bunching. 

We begin by putting an odd cat state $\ket{\mathrm{cat}_-}$ with $t=2$ terms into the first $m=2$ modes. Then Eq.~\ref{eq:sub} becomes,
\begin{eqnarray} \label{eq:exp}
\gamma_s &\approx& \frac{1}{(2\alpha)^2\sqrt{S_1!S_2!}} \sum_{t_1,t_2=1}^{2} (-1)^{\sigma(\vec{t})} \prod_{j=1}^2 (\beta_{t_1,t_2}^{(j)})^{S_j} \nonumber \\
&=& \frac{1}{(2\alpha)^2\sqrt{S_1!S_2!}} \sum_{t_1,t_2=1}^{2} (-1)^{\sigma(t_1+t_2)} (\beta_{t_1,t_2}^{(1)})^{S_1}(\beta_{t_1,t_2}^{(2)})^{S_2} \nonumber \\
&=&  \frac{1}{(2\alpha)^2\sqrt{S_1!S_2!}} \left[
(\beta_{1,1}^{(1)})^{S_1}(\beta_{1,1}^{(2)})^{S_2}
-(\beta_{1,2}^{(1)})^{S_1}(\beta_{1,2}^{(2)})^{S_2} \right.  \nonumber \\
&-&\left. (\beta_{2,1}^{(1)})^{S_1}(\beta_{2,1}^{(2)})^{S_2}
+(\beta_{2,2}^{(1)})^{S_1}(\beta_{2,2}^{(2)})^{S_2}
\right]
\end{eqnarray}

Now to calculate the $\beta_{\vec{t}}^{(j)}$'s for this case we first take the tensor product between the first two modes. Ignoring the normalization factor this yields, 
\begin{eqnarray} 
\ket{\mathrm{cat}\_}&=&(\ket{\alpha}-\ket{-\alpha})\otimes(\ket{\alpha}-\ket{-\alpha}) \nonumber \\
&=&\ket{\alpha,\alpha}-\ket{\alpha,-\alpha}-\ket{-\alpha,\alpha}+\ket{-\alpha,-\alpha}. \nonumber \\
\end{eqnarray}

Next we pass them through a 50/50 beamsplitter,
\begin{equation}
U\ket{\mathrm{cat}\_}=\ket{\sqrt{2}\alpha,0}-\ket{0,\sqrt{2}\alpha}-\ket{0,-\sqrt{2}\alpha}+\ket{-\sqrt{2}\alpha,0}.
\end{equation}
Now we read off the $\beta_{\vec{t}}^{(j)}$'s to be
\begin{eqnarray}
\beta_{1,1}^{(1)}&=&\beta_{1,2}^{(2)}=\sqrt{2}\alpha \nonumber \\
\beta_{2,2}^{(1)}&=& \beta_{2,1}^{(2)}=-\sqrt{2}\alpha \nonumber \\
\beta_{1,2}^{(1)}&=&\beta_{2,1}^{(1)}=\beta_{1,1}^{(2)}=\beta_{2,2}^{(2)}=0.
\end{eqnarray}

Now Eq.~\ref{eq:exp} becomes,
\begin{eqnarray} \label{eq:b}
\gamma_s &=&  \frac{1}{(2\alpha)^2\sqrt{S_1!S_2!}} \left[
(\sqrt{2\alpha})^{S_1}(0)^{S_2}
-(0)^{S_1}(\sqrt{2}\alpha)^{S_2} \right. \nonumber \\
&-& \left. (0)^{S_1}(-\sqrt{2}\alpha)^{S_2}
+(-\sqrt{2}\alpha)^{S_1}(0)^{S_2}
\right].
\end{eqnarray}

Because we are dealing in the limit of small $\alpha$, a non-zero number arbitrarily close to zero raised to a zero power is one, so the terms $0^{S_j}=\delta_{S_j,0}$. Now Eq.~\ref{eq:b} becomes,
\begin{eqnarray} \label{eq:Z}
\gamma_s &=&  \frac{1}{(2\alpha)^2\sqrt{S_1!S_2!}} \left[
(\sqrt{2\alpha})^{S_1}\delta_{S_2,0}
-(\sqrt{2}\alpha)^{S_2}\delta_{S_1,0} \right. \nonumber \\
&-& \left. (-\sqrt{2}\alpha)^{S_2}\delta_{S_1,0}
+(-\sqrt{2}\alpha)^{S_1}\delta_{S_2,0}
\right].
\end{eqnarray}

For this example we know that there are three possible signature outcomes. We expect that the configuration $S_1=S_2=1$ is not possible due to HOM photon bunching and thus in this case $\gamma_s=0$. For configurations $S_1=0$ and $S_2=2$ or $S_1=2$ and $S_2=0$ we would expect a non-zero configuration amplitude of $\gamma_s=1/2$ in each case. Next, we will show that this is indeed the case. 
\subsection{Configuration $S_1=S_2=1$}
With configuration $S_1=S_2=1$ Eq.~\ref{eq:Z} becomes,
\begin{eqnarray}
\gamma_s &\approx& \frac{1}{4\alpha^2} \left[ 
(\sqrt{2}\alpha)\delta_{1,0}
-(\sqrt{2}\alpha)\delta_{1,0} \right. \nonumber \\
&-&\left.(-\sqrt{2}\alpha)\delta_{1,0}
+(-\sqrt{2}\alpha)\delta_{1,0} \right] \nonumber \\
&=& 0,
\end{eqnarray}
which vanishes as expected. 

\subsection{Configuration $S_1=0$ and $S_2=2$}
With configuration $S_1=0$ and $S_2=2$ Eq.~\ref{eq:Z} becomes,
\begin{eqnarray} \label{}
\gamma_s &=&  \frac{1}{(2\alpha)^2\sqrt{0!2!}} \left[
(\sqrt{2\alpha})^{0}\delta_{2,0}
-(\sqrt{2}\alpha)^{2}\delta_{0,0} \right. \nonumber \\
&-& \left. (-\sqrt{2}\alpha)^{2}\delta_{0,0}
+(-\sqrt{2}\alpha)^{0}\delta_{2,0}
\right] \nonumber \\
&=& \frac{1}{4\alpha^2\sqrt{2}} \left[
-2\alpha^{2}
-2\alpha^{2} 
\right] \nonumber \\
&=& -\frac{1}{\sqrt{2}},
\end{eqnarray}
and the corresponding classical probability is $1/2$ as expected.

\subsection{Configuration $S_1=2$ and $S_2=0$}
With configuration $S_1=2$ and $S_2=0$ Eq.~\ref{eq:Z} becomes,
\begin{eqnarray} \label{}
\gamma_s &=&  \frac{1}{(2\alpha)^2\sqrt{2!0!}} \left[
(\sqrt{2\alpha})^{2}\delta_{0,0}
-(\sqrt{2}\alpha)^{0}\delta_{2,0} \right. \nonumber \\
&-& \left.(-\sqrt{2}\alpha)^{0}\delta_{2,0}
+(-\sqrt{2}\alpha)^{2}\delta_{0,0}
\right] \nonumber \\
&=& \frac{1}{4\alpha^2\sqrt{2}} \left[
2\alpha^{2}
+2\alpha^{2}
\right] \nonumber \\
&=& \frac{1}{\sqrt{2}},
\end{eqnarray}
again with classical probability $1/2$ as expected.

Thus, our result generalizes to the expected results for passing a single photon Fock state inputted in modes one and two through a Hadamard gate. This shows that our cat state generalization works for the odd cat state in the limit of small $\alpha$, which is equivalent to Aaronson \& Arkhipov's boson-sampling. 

%
%

\section{Non-zero amplitude odd cat states as an error model} \label{app:poly_bound}

According to the error bound derived by Aaronson \& Arkhipov, the probability of sampling from the correct distribution must not exceed the bound of $1/\mathrm{poly}(n)$ in order for it to implement  classically hard boson-sampling. The correct input distribution is $\ket{1,\dots,1,0,\dots,0}$ and the probability of successfully sampling from it depends on the odd cat states since we input odd cat states in every mode requiring a $\ket{1}$ and vacuum in the remaining modes. Thus, the single-photon component of the odd cat state must be successfully sampled $n$ times. 

Consider the odd cat state from Eq.~\ref{eq:app_odd_cate_state}. The amplitude of the single photon term of an odd cat state is given by,
\begin{equation}
\gamma_1 = \frac{f_1(\alpha) - f_1(-\alpha)}{\sqrt{2(1-e^{-2|\alpha|^2})}},
\end{equation}
where $f_n(\alpha)$ is defined in Eq.~\ref{eq:fna}. In order to get the classical probability we calculate $\gamma_1^2.$ The amplitude of the $n=1$ coherent state photon term is then,
\begin{eqnarray} \label{eq:fna1}
f_1(\alpha) &=& \alpha e^{-\frac{|\alpha|^2}{2}}, 
\end{eqnarray}
thus, the probability of having sampled from the correct term is,
\begin{eqnarray}
P &=& {\gamma_1}^{2n} \nonumber \\
&=& \left(\frac{\alpha e^{\frac{-|\alpha|^2}{2}} - (-\alpha)e^{\frac{-|\alpha|^2}{2}}}{\sqrt{2(1-e^{-2|\alpha|^2})}}\right)^{2n} \nonumber \\
&=& \left(\frac{2\alpha e^{\frac{-|\alpha|^2}{2}}}{\sqrt{2(1-e^{-2|\alpha|^2})}}\right)^{2n}\nonumber \\
&=&  \left(\frac{4\alpha^2 e^{-|\alpha|^2}}{2(1-e^{-2|\alpha|^2})}\right)^{n}\nonumber \\
&=& \left(\frac{2\alpha^2}{e^{|\alpha|^2}(1-e^{-2|\alpha|^2})}\right)^{n}\nonumber \\
&=& \left(\frac{2\alpha^2}{e^{|\alpha|^2}-e^{-|\alpha|^2}}\right)^{n}\nonumber \\
&=& \left(\alpha^2\mathrm{csch}(|\alpha|^2)\right)^{n}\nonumber \\
&=&\alpha^{2n} \mathrm{csch}^n(|\alpha|^2),
\end{eqnarray}
where the hyperbolic trigonometric identity $\mathrm{csch}(x)=2/(e^x-e^{-x})$ was used.
Following A\&A's given bound this requires that,
\begin{equation}
\alpha^{2n} \mathrm{csch}^n(|\alpha|^2) > 1/\mathrm{poly}(n),
\end{equation}
in order for the sampling problem to be in a regime which is provably computationally hard.

\end{document}